# Space Weathering on Near-Earth Objects investigated by neutral-particle detection

C. Plainaki[(1)], A. Milillo[(1)], S. Orsini[(1)], A. Mura[(1)], E. De Angelis[(1)], A. M. Di Lellis[(2)], E. Dotto[(3)], S. Livi[(4)], V. Mangano[(1)], S. Massetti[(1)], M. E. Palumbo[(5)]

[(1)] INAF - Istituto di Fisica dello Spazio Interplanetario Via del Fosso del Cavaliere, 00133 Roma, Italy, christina.plainaki@ifsi-roma.inaf.it

[(2)] AMDL srl, Rome, Italy, amdlspace@gmail.com

[(3)] INAF - Osservatorio Astronomico di Roma, Monteporzio, Italy, dotto@mporzio.astro.it

[(4)] SwRI, San Antonio, USA, Stefano.Livi@swri.edu

[(5)] INAF - Osservatorio Astrofisico di Catania, Italy, mepalumbo@oact.inaf.it

## Abstract

The ion-sputtering (IS) process is active in many planetary environments in the Solar System where plasma precipitates directly on the surface (for instance, Mercury, Moon, Europa). In particular, solar-wind sputtering is one of the most important agents for the surface erosion of a Near-Earth Object (NEO), acting together with other surface release processes, such as Photon Stimulated Desorption (PSD), Thermal Desorption (TD) and Micrometeoroid Impact Vaporization (MIV). The energy distribution of the IS-released neutrals peaks at a few eVs and extends up to hundreds of eVs. Since all other release processes produce particles of lower energies, the presence of neutral atoms in the energy range above 10 eV and below a few keVs (Sputtered High-Energy Atoms - SHEA) identifies the IS process. SHEA easily escape from the NEO, due to NEO's extremely weak gravity. Detection and analysis of SHEA will give important information on surface-loss processes as well as on surface elemental composition. The investigation of the active release processes, as a function of the external conditions and the NEO surface properties, is crucial for obtaining a clear view of the body's present loss rate as well as for getting clues on its evolution, which depends significantly on space weather.

In this work, an attempt to analyze the processes that take place on the surface of these small airless bodies, as a result of their exposure to the space environment, has been realized. For this reason a new space weathering model (Space Weathering on NEO - SPAWN), is presented. Moreover, an instrument concept of a neutral-particle analyzer specifically designed for the measurement of neutral density and the detection of SHEA from a NEO is proposed.

## 1. Introduction

The Near-Earth Objects (NEOs) are asteroids and comet nuclei in an evolving population with a lifetime limited to a few million years, having orbits with perihelion distances <1.3 A.U.. Hence, they periodically approach or intersect the orbit of the Earth (Lazzarin et al., 2004a). The diversity among these objects is emphasized by the variety of different taxonomic types existing within a specific population. Currently, two sources for NEOs have been identified. The principal one is the main asteroid belt, where gravitational perturbations by the giant planets and Mars cause dynamical resonances which provide escape routes (Lazzarin et al., 2004b). The second source is represented by extinct comets. In fact, a certain number of NEOs may represent the final evolutionary state of comets, that is, a de-volatilized nucleus (Harris and Bailey, 1998). NEOs, being representatives of the population of asteroids and dead comets, are thought to be the primitive leftover building blocks of the Solar System formation process offering clues to the chemical mixture from which the planets were formed. Although good spectral matches among some NEOs and meteorite types have been found (e.g. Lazzarin et al. 2004a), the link between NEOs and meteorites is not completely understood, constituting, therefore, an intriguing issue which is currently under continuous research.

### 1.1. NEO characterization problems

Investigation of the NEO erosion and evolution can be realized via the identification and the localization of the consequences of the so called space weathering, i.e. the physical processes taking place on the surface of the body as a result to its exposure to the space environment. Such studies could be performed by any NEO dedicated space platform, in the frame of the next international programs, by observing the gas species expanding from the asteroid surface. It is already found that space weathering influences some categories of asteroids (Ueda et al., 2002). The exact mechanism is still under discussion. Mostly the difficulties are in inferring mineral information from spectral data and in identifying effects of space weathering by comparison between asteroids and laboratory (or meteoritic) spectra (see however Strazzulla et al. 2005 and references therein). It should be pointed out that, until present, the principal method to classify NEOs taxonomically and to address their origin, was their spectroscopic characterization (Tholen, 1984, 1989; Tholen and Barucci, 1989). Remote sensing has been an essential approach to learning about the nature of asteroids, while spacecraft could visit only very few of them. Consequently, various studies on identifying the surface composition of these objects and the effects of space weathering on their surface have been realized, using observational data in the visible and near infra-red region (Lazzarin et al., 2004a; 2004b). In specific, Lazzarin et al. (2004a) searched for ordinary chondrite parent bodies among NEOs, as well as for aqueous altered materials, performing comparison of the obtained spectra with the largest sample of the main belt asteroids spectra available in the literature (Bus, 1999; Bus and Binzel, 2002a; 2002b). However, the inferred classification was obtained taking into account the mean spectral classes, and therefore loosing the information contained in all single spectra (Lazzarin et al., 2004a). As a consequence, one should conclude that although the studies based on spectral analysis provide useful information, however, they are quite limited and, therefore, direct in situ observations have to be performed in order to improve the situation.

On the other hand, studying the effects of space weathering on NEOs on the basis of laboratory experiments constitutes a task of significant difficulties. Important efforts on modelling the asteroids space weathering on the basis of lab experiments have been realized, nevertheless, the conditions during these experiments are more or less focused, succeeding in reproducing only a fraction of the real situation in space. Hapke (1968) interpreted telescopic data on the optical properties of the Moon in terms of equivalent laboratory measurements of powders of terrestrial rocks and meteorites. Anticipating that the solar wind might space-weather the lunar surface, compared his lab measurements of the powders irradiated with a beam of 2 keV hydrogen ions (to simulate the solar wind) with measurements of un-irradiated powders; Hapke found that the irradiation greatly modified the optical properties of the sample, concluding that the lunar surface might consist of basalts and not ordinary chondrites, an hypothesis that turned out to be correct. Nevertheless, his experimental results (as well as those of others at that time) were clouded by issues such as contamination, which continued to plague laboratory simulations of solar-wind bombardment through the 1970s (Chapman, 2004). Similar experimental efforts attempting to reveal the features of space weathering, lead to relevant space weathering in debates (Chapman, 1996). For example, Chapman and Salisbury (1973) while trying to match asteroid spectra with laboratory spectra of meteorite powders, they noticed both similarities between S-types and Ordinary Chondrites (OC) (e.g., absorption band near 0.95 μm) and differences (the straightened, reddish slope of S-type spectra through the visible and the diminished depth of the absorption band). In the analytical interpretation of their results, they noted certain difficulties, regarding (*a*) the uncertainty whether vitrification of OC material would behave like lunar vitrification, (*b*) the impact velocities in the asteroid belt that might be too low for efficient vitrification, (*c*) the expected immaturity of asteroidal regoliths owing to low gravity and hence loss of most ejecta to space rather than reincorporation into the regolith, and most important (*d*) the apparent lack of space weathering on a body of lunar-like, basaltic composition (e.g. Vesta) (Chapman, 2004). Similar issues were later raised by Matson et al. (1976; 1977), again in the context of the then-accepted paradigm for lunar space weathering. Moreover, recently Ueda et al. (2002) has shown how the space weathering and the grain size effects influence the NEO classification. Nevertheless, their results indicate the difficulties in inferring mineral information from spectral data, as well as in identifying the effects of space weathering by comparison between asteroids and laboratory (or meteoric) spectra (Lazzarin et al., 2004a).

In summary, one should consider that the real situation is even more complex than that modelled on the basis of either visible and IR observations or lab experiments, which, in any case, present considerable difficulties although they have been studied in great detail using returned lunar samples. Moreover, the composition and space environment of these samples are quite different from those of asteroids. Therefore interpretations based explicitly on the above methods cannot always be considered definitive. Consequently, direct in situ observations can help significantly in providing quantitative information on asteroid space weathering, contributing essentially in extending our current knowledge on this issue.

### 1.2. Neutral-particle release processes

Due to space weathering, NEOs suffer erosion and surface alteration. The most important processes taking place on their surface are the following: a) solar-wind particles interaction with the surface atoms, b) solar and galactic cosmic-ray bombardment, c) solar photon irradiation and d) micrometeorites precipitation. The relevant surface release processes at these distances from the Sun are Ion Sputtering (IS), Photon Stimulated Desorption (PSD), Thermal Desorption (TD) and Micrometeoroid Impact Vaporization (MIV). The IS process, active in many planetary environments in the Solar System -for instance, Mercury (e.g., Milillo et al., 2005), Moon (e.g., Wurz et al., 2007), Europa (e.g., Eviatar et al., 1985)-, is defined as the removal of a part of atoms or molecules from a solid surface, due to the interaction of a projectile ion with target electrons and nuclei, as well as secondary cascades of collisions between target atoms (Sigmund, 1981). It is one of the most important processes, the products of which depend on the composition and the chemical structure of the surface. Although research on the release of sputtered particles from solids is being realized for more than 40 years, the phenomenon is yet not fully understood. The PSD refers to the desorption of neutrals or ions as a result of direct excitation of surface atoms by photons (Hurych, 1988), whereas the TD exists when the thermal energy of an atom exceeds the surface binding energy. The MIV refers to the impact vaporization caused by micrometeorites hitting the surface of an asteroid. Since the other surface release processes are mainly active in the dayside, MIV is likely to be the most important process for the nightside of airless bodies (Killen and Ip, 1999).

Identifying the rate of surface aging and space weathering, due to all types of release-processes, is of particular importance for the airless NEOs. However, different release processes produce particles within different energy ranges (Wurz and Lammer, 2003). TD and PSD are more effective for volatiles (like H, He, Na, K, S, Ar, OH) and have typical energy lower than 1 eV. IS and MIV are relatively stoichiometric processes and are effective also for refractory species. Nevertheless, the particles produced via MIV have a Maxwellian distribution with a peak value expected to be in the range 2500–5000K (Eichhorn, 1978) that corresponds to a peak energy for the emerging particles of ~0.6 eV. The IS is the most energetic release process among all others, being capable of releasing particles at energies above 10 eV (Milillo et al., 2005). The possibility to discriminate the contribution of different release processes will permit to speculate on the surface erosion under different environmental conditions. Consequently, solar-wind sputtering investigation provides important clues on the history evolution of a planetary body. The 3-D flow velocity of the ion-sputtered neutrals detected through high spatial resolution measurements in conjunction with solar-wind analyzers will allow the exact determination of the surface area from which the particles have been released; hence, it will be possible to map the location of the sputtering process on the surface and also to image the surface loss rate.

Consequently, laboratory and theoretical simulations jointly with in situ observations of the release products are crucial tools for further investigation. A study on asteroids exosphere based on the simulation of the various release processes that take place on the surface of the body has been performed by Schläppi et al. (2008), for the asteroids (2867) Steins and (21) Lutetia, in preparation of the Rosetta flybys. They found that the solar-wind sputtering is the most important exospheric supply process on the sunlit side of an asteroid. It should be underlined that IS is the most effective process in releasing the higher energies particles as well as refractory species (e.g., Si, Al, Mg; see Table-I). This means that it can be discriminated from all other release processes, in case that high-energy particles are being observed. On the basis of this view, we define as Sputtered High Energy Atoms (SHEA) the sputtered particles that have energies in the range between 10 eV and a few keV. We concentrate on the analysis of SHEA, because they have energies inside an energy window in which particles released from all other processes are negligible.

In our study we focus on the IS process taking place on the surface of a NEO, presenting a new model for describing the emerging particle flux and density distribution (Section 2). The results of our calculations for three different cases of NEO surface compositions considered are presented in Section 3. Moreover, the feasibility of monitoring the SHEA emitted from the NEO surface is investigated. An instrument concept of a neutral-particle analyzer, specifically designed for the detection of neutrals released from a NEO, is presented in Section 4.

## 2. The SPAWN model

The SPAce Weathering on NEO (SPAWN) model is a new model intended to study space weathering effects taking place on the surface of an asteroid in the near-Earth interplanetary environment. In its current version, this model assumes three different cases of NEO composition: a) composition similar to a CI chondrite type asteroid, b) composition similar to a CM chondrite type asteroid and c) composition similar to Tagish Lake asteroid (D-type). The exact chemical compositions used in this analysis were originally taken from Brown et al. (2000) and then they were transformed to atoms % unit (Table-I). Carbonaceous asteroids are of particular interest due to the fact that carbonaceous meteorites on Earth are the best preserved witnesses of the early solar system formation. In general, a carbonaceous chondrite or a C-type chondrite contains high levels of water and organic compounds whereas its bulk composition is mainly silicates, oxides and sulfides. In specific, CI chondrites contain small chondrules (typically 0.1 to 0.3 mm in diameter) and similar-sized refractory inclusions. Among the various processes taking place on the NEO surface (see Section 1), the SPAWN model takes into consideration the IS process as well as the PSD and TD processes. However, the most important exospheric process on the sunlit side of an asteroid at distances above 2 AU is the IS (Scläppi et al., 2008). Moreover, in the SPAWN model we have assumed that the deflection of the solar-wind ions by possible intrinsic magnetic fields possessed by the NEO (Richter et al., 2001) is negligible and that sputtering occurs only on the dayside of the NEO. Nevertheless, an estimation of PSD and TD generated exosphere is given in Section 3 for comparison.

The products of IS depend on the composition and the chemical structure of the surface. The energy transferred in the collision between an energetic ion and a surface atom, $T$, is given by (Hofer, 1991):

$$T = T_m \cos^2 \theta_r \tag{1}$$

where $T_m$ is the maximum energy transferred, given by:

$$T_m = \frac{4m_1 m_2}{(m_1 + m_2)^2} E_i \qquad (2)$$

where $E_i$ is the energy of the incident particle, $m_1$ is its mass, $m_2$ is the mass of the surface target particle released and $\theta_r$ is the recoil angle, defined as the angle between the surface normal and the velocity direction of a target particle, after its collision either with the primary projectile (primary collision) or with a sputtered substrate atom (secondary collision). In Fig. 1 the geometry of the model is presented. The incident angle of the projectile ion, $\theta_i$, is defined as the angle between its velocity direction and the surface normal, whereas the ejection angle, $\theta_e$, is defined as the angle between the velocity direction of a target atom escaping from the surface and the surface normal. From its definition, it is clearly seen that $\theta_r$ regards collisions that can take place at any depth or at the surface and depends on the material structure itself as well as on the incident particle energy. During a sputtering cascade, the recoil angle of the last collision between two particles and the ejection angle are identical. The main scope of this study is the revealing of the sputtered atoms distribution at a region above the NEO surface. For this reason, we take into account only the angle at which the particles finally abandon the surface, and not the details of the sputtering cascade inside the body substrate.

The incident angle at which an ion projectile impinges the NEO surface is not always normal to the surface. However, in the regime of the cascade sputtering, the ejection distribution is rotationally symmetric with respect to the surface normal, whatever the direction of the incoming beam (Hofer, 1991). This is a consequence of the recoil cascade itself, that has forgotten the details of the primary projectile target interaction while slowing down (Behrisch, 1964; Kaminsky, 1965). Finally, conservation of momentum is taken into account in this model.

The surface binding forces influence significantly the energy spectrum of the sputtered particles. Depending on the type of surface barrier potential the intersection of the cascade by the surface either causes a small modification of the ejection spectrum or dominates the whole low-energy part ($\leq 50\, eV$). The charge state of the released particles can be neutral or ionized. While the neutral atoms expand radially from the NEO, the ions are loaded in the solar wind. Anyway, the ionized component is less than 1% of the neutral one, depending on the element, the surface mineralogy and the impacting ion (Benninghoven et al. 1987); hence, it is negligible as a first approximation. Consequently, in this study we focus on the neutral releasing. The energy distribution of the ejected neutral atoms affords vital information about the sputtering phenomenon as a whole (Fig. 2) and it can be expressed as (Sigmund 1969 ; Thompson 1968):

$$f_s(E_e, E_i) = \begin{cases} \sim \dfrac{E_e}{(E_e + E_b)^3}\left[1 - \left(\dfrac{E_e + E_b}{T_m}\right)^{1/2}\right] & E_e \leq E_i - E_b \\ 0 & E_e > E_i - E_b \end{cases} \tag{3}$$

where $E_e$ is the ejection energy and $E_b$ is the surface binding energy. The shape of the energy distribution function for an incident hydrogen particle of energy ~1000 eV for different binding energies is shown in Fig.2. The surface binding energy, corresponding to $E_b$, in relation (3), influences significantly the energy spectrum of the sputtered particles (e.g., Mura et al., 2005). The spectrum determined experimentally has a distinct maximum. On the basis of equation (3) it can be easily shown that for impinging particle energies much greater than the binding energy, the spectrum peaks at $E_b / 2$. Moreover, one should take into account that the model of a spherically symmetric surface barrier is generally considered inappropriate at least as far as the total emission is considered (Hofer, 1991). However, as a first approach, in this analysis, we have used a binding energy value for the whole NEO surface, equal to 2 eV. The SHEA are about 1% of the total (for 1 keV impinging ion and 2 eV binding energy). The present simulations consider an average energy of solar wind, that is 1 keV protons, nevertheless, the proton energies can reach up to 4 keV in fast solar-wind cases. In this cases, the energy distribution of the sputtered particles can extended at above 1 keV.

The sputtering effect taking place on a NEO surface was studied via the SPAWN model applying the Monte Carlo (MC) Method for a number of test particles equal to 100000, in order to calculate the emerging sputtered neutral-particle flux and density over a cubic space of 3km x 3km x 3 km around the NEO. The Monte Carlo code assumes angular and velocity distributions for the sputtering process and then tracks the trajectory of each released particle through the exosphere. The SPAWN code uses a fixed step $dx = 2R_{NEO} / n$, where $R_{NEO}$ is the NEO radius and $n$ is the number of bins, in order to derive the flight time, $dt$, for each step (i.e. $dt = \frac{dx}{\upsilon}$, where $\upsilon$ is the particle's velocity ). A solar-wind proton flux at 1 AU of $10^{12}$ $m^{-2}s^{-1}$ was considered as the total amount of the impinging particles, whereas different species of ejected particles were considered (see Table-I). The NEO radius was supposed to be 0.5 km; its mass was taken as $10^{12}$ kg. The yield, which is the number of sputtered atoms produced by one single impinging ion, strongly depends on the composition and the chemical structure of the surface as well as on the mass and energy of the impinging ions. For example surfaces slightly porous result in reduced sputtering yields due to both shadowing and sticking effects of ejected particles from one grain onto another (Jurac et al., 2001). The influence of the type of NEO surface on the total flux ejected is included in the yields (Hapke and Cassidy, 1978). Therefore, the SPAWN model, in its current version, has assumed yields based on the regolith particle sputtering coefficients of elements given by Starukhina (2003). These coefficients are normalized by their concentrations and by the total sputtering coefficient, and they refer to the case that the target contains implanted H. As a consequence, for the common species between NEOs and the Moon (i.e. H, Mg, Al, Si, Ca and Fe), we calculated the sputtering yields by multiplying the yield average value of 0.05 with the respective coefficients given by Starukhina (2003). For the rest of the elements we assumed the sputtering yield of 0.05. In the future, a more accurate choice is intended to be performed for these species, taking also into consideration that binding energies influence the yield. However, simulations have shown that variations of the binding energy within the range 1 – 3 eV, which is typical of solid material, change the sputtering yield no more than 2% of their average values at $E_b = 2\,\mathrm{eV}$ (Starukhina, 2003). Thus, we do not expect that a more accurate choice of the binding energies would change significantly the present results.

For the polar ejection angle, $\theta_e$, a $\cos^k(\theta_e)$ function can be used in general for sputtering (Schläppi et al., 2008), where the exponent $k$ depends on the structure of the surface and equals to 1 for a porous regolith (Cassidy and Johnson, 2005).

A summary of the input parameters used in the SPAWN model, is presented in Table-II.

## 3. Results of the simulations - Discussion

Fig. 3 (left panel) presents the simulated intensity of the total sputtered flux produced by all species of sputtered particles emerging from a NEO surface consisting of CI type chondrites. This flux is integrated over all energies and over all emission directions. The energy of the impinging protons for all cases was considered as 1000 eV. Solar-wind energies can be even higher (up to 4 keV) thus also the sputtered high energy tail could extend at higher energies. The color bar scale on the right gives the decimal logarithm of the sputtered particle flux (in particles $m^{-2}$ $s^{-1}$). It is clearly seen that up to an altitude of about 1 km above the NEO surface, the intensity of the sputtered-particle flux reaches the value of $10^{11}$ particles $m^{-2}$ $s^{-1}$. The higher energy (>10eV) particle flux derived by previous considerations results in about $10^{9}$ particles $m^{-2}$ $s^{-1}$. On the basis of the above value of the emerging neutral-particle flux, the maximum global release rate from the NEO can be estimated. For a NEO radius of 0.5 km the maximum global release rate of sputtered neutral atoms is about $3.14 \cdot 10^{17}$ particles/s. Schläppi et al. (2008) have calculated for the asteroid (2867) Steins (radius ~ 4.6 km, distance from the Sun ~ 2.364 AU) a sputtering rate of $6.84 \cdot 10^{17}$ particles/s, for the emerging oxygen atoms. Similar escape rates for different asteroid sizes can be explained since our simulation mostly considers hydrogen atoms, while Schläpppi et al. (2008) consider only oxygen particles, which have the highest relative composition on the surface of asteroid Steins.

Therefore, although the asteroid Steins is bigger in dimensions than the NEO studied, the different surface characteristics result in similar sputtered particle rates for the two cases.

In Fig. 3 (right panel) it is shown that the simulated intensity of the total density, produced by all species of sputtered particles emerging from a NEO surface consisting of CI type chondrites, is ~ $3 \cdot 10^6$ particles $m^{-3}$ near the NEO surface. This result is in good agreement with the calculations made by Schläppi et al. (2008) according to which, the exobase density of neutral particles emerging from asteroid Lutetia is about $6 \cdot 10^6$ particles $m^{-3}$.

Apart from IS, PSD and TD contribute to the total sputtered particle density emerging from the NEO surface. In our model, the PSD release process is applied only to the H-and-C NEO populations, since it constitutes a process that regards exclusively volatile species. We perform an estimation of the PSD contribution, assuming the energy distribution given by Wurz and Lammer (2003), a photon flux in the orbit of Earth equal to $3.31 \cdot 10^{19}$ $m^{-2}s^{-1}$ (Schläppi et al, 2008), a PSD cross section equal to $10^{-24}$ $m^2$ for the volatile components (i.e. H and C, with the relative compositions given in Table I), an asteroid regolith surface density of $7.5 \cdot 10^{18}$ $m^{-2}$ (Wurz and Lammer, 2003). The results of our simulations show that the total density of the volatiles emerging from the NEO surface, via the PSD process, is ~ $1 \cdot 10^8$ particles/$m^3$. Moreover, we have realized a rough estimation of the particles emerging via TD for different values of the surface temperature *T*. The released volatiles particle density, emerging from TD, varies from ~$10^4$ particles/$m^3$ (for *T*=400K) to ~$5 \cdot 10^8$ particles/$m^3$ (for *T*=500K). Summarizing the results from simulating both PSD and TD , we find that the total released particle density varies from ~ $1 \cdot 10^8$ particles/$m^3$ to ~ $6 \cdot 10^8$ particles/$m^3$. This value is similar to that calculated by Schläppi et al (2008) for asteroid Steins (~ $2 \cdot 10^8$ particles/$m^3$), being a little higher. However one should take into account that since asteroid Lutetia is further away from the Sun, at a distance of about 2.72 A.U, this small difference in the expected PSD and TD released particle density is reasonable. On the basis of the above mentioned estimations, the fluxes emerging via the processes of PSD and TD are 1.5-2 orders of magnitude more intense than those emerging via solar-wind sputtering. On the contrary, as far as the MIV is concerned, Schläppi et al. (2008) have found that for the asteroids Steins and Lutetia (at 2.14 and 2.72 AU respectively) the solar-wind sputtering contribution is about one order of magnitude higher than impact vaporization at the subsolar point.

Fig. 4 presents the simulated sputtered particle density of the two most abundant species emerging from a NEO surface consisting of CI type chondrites (i.e. H and Mg, with respect to Table-I). It is clearly seen that the expected H density is bigger than that of Mg by a factor of ~10. This difference is related to the higher H abundance (see Table-I) and to the higher H sputtering yield (Starukhina, 2003). Moreover, from the comparison between Figures 3 and 4 it is derived that the sputtered-H density constitutes ~ 90% of the total one.

On the basis of the results on the emerging neutral-particle fluxes stated above, the NEO erosion rate due to solar-wind sputtering can be estimated. If the flux of the sputtered atoms leaving the NEO surface is equal to $j= 10^{11}$ particles $m^{-2}$ $s^{-1}$ then the erosion rate $j\cdot\omega$ is 0.3 Å/year, where $\omega=10^{-29}$ $m^3$ is the atomic volume in a solid substance. This result is similar to the estimation performed for the lunar surface, 0.2 Å/year, in case of solar-wind sputtering (Starukhina, 2003). Taking into consideration that the upper limit of the rate of erosion under micrometeoritic bombardment is calculated to be 0.13 Å/year (Starukhina, 2003; Lebedinets, 1981), the contribution of sputtering to irreversible erosion possibly exceeds that of micrometeoritic bombardment. On the other hand, the erosion rate under PSD is estimated at ~10 Å/year, if the corresponding neutral release rate is considered to have been constant. Nevertheless, since in the past NEO was located farther from the Sun and its temperature was lower, the erosion due to TD must have been negligible. Moreover, it should be underlined that PSD works only for volatiles (i.e. H and C, in the case of the NEO being studied here). All other elements cannot be released from the surface via PSD and, thus, once H and C are removed the process comes to a standstill until material diffuses from the interior to the surface or fresh material is exposed by other processes (e.g. sputtering). However, it should be mentioned that a detailed study on the global erosion of a NEO, demands, a Sun evolving model, that will include temporal changes of temperature, luminosity and solar wind as well.

The simulated intensity of the total sputtered flux, emerging from a NEO surface consisting of CM type chondrites is presented in Fig. 5. No big differences exist between the left panel of Fig. 3 and Fig. 5. Also in the case of CM chondrite-type NEO surface, significantly big values of neutral sputtered particles fluxes (up to $10^{11}$ particles $m^{-2}$ $s^{-1}$) appear in the region near the NEO surface, and in particular in a region extending from the NEO surface up to an altitude of about 0.75 km above that point. Therefore, the region of significant emitted neutral-particle flux seems to be less extended than that in case of CI chondrite type NEO surface.

In Fig. 6 the sputtered-H density difference between the cases of CI and CM chondrite-type NEO surfaces is presented. One can see that, according to the simulations, a flux difference of ~1 order of magnitude smaller than the actual densities is expected. Moreover, the sputtered-H density expected to be measured, in case the NEO composition is similar to that of a CI chondrite, is slightly bigger than that in case of a CM chondrite one. The biggest difference exists in the regions near the NEO surface.

The estimated total flux and density of all sputtered particles emerging from a NEO surface with a composition similar to that of D-type asteroid Tagish Lake are similar to those calculated in the previous cases (CI and CM chondrite type compositions). Also in this case significantly big values of neutral sputtered particles fluxes (up to $10^{11}$ particles $m^{-2}$ $s^{-1}$) appear in the region near the NEO surface, and in particular in a region extending from the NEO surface up to an altitude of about 1 km above that point.

Last but not least, it should be mentioned that the simulation considers an average solar-wind condition. In case of solar extreme event activity a greater released sputtered flux is expected.

## 4. A neutral-particle sensor for NEO released signal detection

The investigation of the active processes as a function of external conditions and surface properties is crucial in order to get a clear view of the present loss rate and the evolution of the body, which depends significantly on space weather. To distinguish the active processes, the expanding gas intensity, composition, velocity and direction should be measured.

The bulk of the surface released particles is below the eV range and hence, the measurement of the gas density will provide information about the intensity and the mass of emitted material. The identification of the mass of the released particles will give some hints on the NEO surface composition.

Nevertheless, different release processes are active on the NEO surface and, when only gas density is measured, it is difficult, if not impossible, to discriminate their different contributions or to reconstruct the emission regions on the surface (and derive areas with different release efficiencies). In fact, it is not possible to investigate whether the IS process is active without a measurement of the energy (or velocity) spectra, since this is the only process that releases particles at energies above 10 eV.

Moreover, a good angular resolution is important to identify the regions more active in releasing SHEA, thus evidencing possible anisotropies of solar-wind sputtering and/or of surface properties.

A comparison between present ion fluxes measured at the vicinity of the Earth and SHEA fluxes emitted from a NEO will provide an indication of the regolith efficiency in releasing material when exposed to solar wind.

To detect and to characterize neutral atoms in the energy range of interest, < 1 eV – 1.0 keV, in an environment of photon, electron and ion fluxes, it is required: 1) highly effective suppression of photons, electrons and ions and 2) two sensors for particles above and below 10 eV. In order to optimize the resources the two sensors should be included in a single instrument design. A possible example of instrument concept able to accomplish the described scientific requirements is RAMON (Released Atoms MONitor). This instrument is a new design of a neutral-particle analyser consisting of two neutral atom sensors able to detect and characterize (in terms of Time of Flight (ToF), mass and direction) the neutral atoms released from the NEO surface. In particular,

- SHEAMON (Sputtered High-Energy Atoms MONitor) will investigate the IS process by detecting SHEA between ~20 eV and ~3 keV and by determining their direction and ToF;
- GASP (GAs SPectrometer) will analyse the mass of the gas density.

Similar detectors are included in the BepiColombo payload (SERENA; Orsini et al. 2008a, 2008b) and are proposed for Solar Orbiter mission.

In Fig. 7, the RAMON basic concept is qualitatively given. The incoming radiation made by neutrals, ions and photons impinges upon an aperture. The ions and electrons are deflected by electrostatic lens (B). The neutral particles pass through an entrance of about 1 $cm^2$ which is divided in two parts for detecting gas density and higher energies (>20 eV).

For low-energy particle detection, the neutral particles pass through a carbon nanotube system (C1) that ionizes the particles (Modi et al., 2003). The ionized particles cross an electronic gate (C2) that provides the START of the ToF (example of such time tagging characterization is given in Brock et al. 2000). Then the particles are accelerated up to more than 1 keV and deflected by an electrostatic system (C3) and are detected by a STOP MCP detector (C4). Electrostatic analyzers are extensively studied in the frame of the CLUSTER/CIS instrument (Di Lellis et al. 1993; Rème et al., 1997). The ToF provides information about mass (since the spread in energy is assumed to be negligible).

For detecting particles between 0.02–1 keV, the neutrals pass through a double grating system (with slits of nanometric dimension) (D1) (Orsini et al. 2008b) that provides photon suppression. A shuttering system allows to move the two gratings one with respect to the other in order to permit the neutrals to enter in the sensor only when the slits are aligned (open gate), which defines the START time. Then the neutrals fly into a ToF chamber and are converted into ions by using the technique of neutral-ion conversion surface (D2) (Wurz, 2000). The ionization efficiency is sufficient at the lowest particle energies and even increases for higher energies (Wurz and Wieser, 2006). When particles impact at the conversion surface electrons are released, even at low impact energies (Wieser et al., 2005). An electrostatic system accelerates the released electrons keeping them well aligned to the original projection to the surface impacting point and pushing them toward the MCP detector, which also has position sensing capability (D3). The MCP will provide the STOP signal for the ToF measurement as well as the angular direction of the velocity of the registered neutral particle. The atom converted in ion by the conversion surface will be accelerated and detected by a MCP (D4) that will provide an additional STOP signal. Moreover, for increasing the geometrical factor, the detector can be used in open-gate mode. In this way the ToF can be identified using as START the first MCP signal. However, the energy resolution will be lower, due to the indetermination in the energy and recoil angle after the impact on the conversion surface.

The spacecraft resources for this instrument are not demanding. In fact, the dimension of the whole RAMON instrument is about 20x20x10 $cm^3$ and its mass will be about 2 kg.

The FOV of the two detection systems is 5°×30°. The higher energy distribution will be analyzed with an angular resolution of 5°x2° (high angular resolution mode) or 5°x5° (low angular resolution mode).

Taking into account the instrument elements, the estimate of the SHEAMON geometrical factor is in the range 4 $10^{-4}$-2 $10^{-5}$ $cm^2$ sr, and the GASP efficiency of about 0.14 (counts/s)/$cm^{-3}$.

These sensor characteristics permit a detection of the estimated particle release. In fact, if the estimated particle flux due to IS from NEO is $10^7$ $cm^{-2}$ $s^{-1}$, more than 1200 counts are estimated in the SHEAMON sensor for 1 minute of integration time. The contribution of IS to the gas density is of the order of 10 $cm^{-3}$ close to the surface, we estimate densities at least one order of magnitude higher, when other release processes are considered (see previous section). In this case, for a 1-minute integration time, about 1000 counts in the GASP sensor are expected.

## 5. Conclusions

The particles released from the NEO surface are essentially lost in space since their escape velocity is very low (i.e. 0.52 m/s for a NEO mass of ~$10^{12}$ kg and a NEO radius of ~0.5 km). Given a specific model for the simulation of the various release processes happening on the surface of a NEO, the efficiency of each particle release processes can be estimated. Moreover, identifying the NEO surface properties and its interactions with the solar wind can provide important information on the space weathering effects at which it has been subjected as well as on the surface and the global evolution history of the body.

The SPAWN Model gives the sputtered neutral atoms distribution around a NEO as a result of its exposure to the solar wind. Moreover an estimation of the release due to PSD and TD has been performed. After having applied this model for three different cases of NEO surface compositions, our major results can be summarized as follows:

- In all three different cases of NEO surface compositions significant sputtered fluxes reaching the maximum value of $10^{11}$ particles $cm^{-2}$ $s^{-1}$ around the NEO are expected. The derived high-energy (>10eV) particle (SHEA) flux results in ~1% of the total one.
- In all three different cases of NEO surface compositions, the simulated intensity of the total sputtered density, produced by all species of sputtered particles emerging from a NEO surface, is calculated to be ~3 $\cdot 10^6$ particles $m^{-3}$ near the NEO surface. The contribution to the total density of the volatiles emerging from the NEO surface, via the PSD process, is ~ $1 \cdot 10^8$ particles/$m^3$. The released volatiles particle density, emerging from TD, varies from ~$10^4$ particles/$m^3$ (for $T$=400K) to ~$5 \cdot 10^8$ particles/$m^3$ (for $T$=500K).
- According to the SPAWN model the sputtering loss mechanism is able to remove approximately $3.14 \times 10^{17}$ particles per second. The NEO erosion rate under sputtering is 0.3 Å/year, a result similar to that estimated by Starukhina (2003) for the lunar surface (0.2 Å/year). On the other hand, the erosion rate under PSD is estimated at ~10 Å/year, if the corresponding neutral release rate is considered to have been constant. As much as the erosion rate due to TD is concerned, one should consider that in the past NEO was located farther from the Sun and its temperature was lower. Thus the erosion due to TD must have been negligible. In order to perform a more detailed study on the global erosion of a NEO, one should consider a Sun evolving model, taking into consideration the temporal changes of the temperature, the luminosity and the solar wind as well.

- The sputtered density of H is expected to be higher than that of other species (by a factor of ~10 in all cases of the considered NEO surface compositions).
- If we consider different NEO compositions, the application of the SPAWN model results in sputtered-particle density values that are quite similar. In fact, the differences in the sputtered-particle densities, between cases of different NEO compositions, are of ~1 order of magnitude smaller than the densities. The biggest difference between the two different composition assumptions exists in the regions near the NEO surface.

As a final remark, it should be stated that the global analysis of the sputtering erosion of the NEO surface will give unique information about the present and the past of the NEO's surface, revealing the mechanism in which the solar wind has interacted with its atoms, in the past millions of years. The nowadays technologies permit to develop instruments able to accomplish the investigation of this process. Consequently, in the frame of a possible future mission to a near-Earth asteroid, for example the Marco Polo Sample Return Mission (Barucci et al., 2008; Dotto et al., 2008), the space weathering study constitutes a crucial and attainable scientific objective.

In the future, an improved SPAWN model, that will include the MIV contribution, detailed analysis of PSD and TD released particle distribution, as well as the influence of the surface volatile depletion due to PSD and TD on the IS results, is intended. Finally, a more detailed study of the evolution of a NEO could be performed by considering a realistic Sun evolution model.

## 6. Acknowledgements

The authors thank the two referees for corroborating this paper with very useful comments and suggestions. The authors also thank the Italian Space Agency for supporting their activities.

## Table captions

**Table-I:** Bulk element abundances for CI-chondrites, CM-chondrites and Tagish Lake type chondrites (adapted from Brown et al., 2000).

**Table-II:** Input Parameters of the SPAWN model.

## Figure Captions

***Fig. 1****: Geometry of the SPAWN model. The incident angle of the impinging ion ,$\theta_i$, together with the recoil angle,$\theta_r$, and the emerging angle, $\theta_e$, of the sputtered atom are presented. This figure refers to primary collisions.*

***Fig. 2:*** *Energy distribution function for different binding energies, characterizing the IS release process, when 1 keV hydrogen ions impact the surface and H atoms are released. Different values of the binding energy correspond to different surface properties.*

***Fig.3:*** *Sputtered-particle total flux (left panel) and density(right panel) for impinging particle of energy ~1keV . The NEO surface is assumed to consist of CI chondrites.*

***Fig.4:****H (left panel) and Mg(right panel) sputtere- particle density for impinging particle of energy ~1keV. The NEO surface is assumed to consist of CI chondrites.*

***Fig.5:*** *Sputtered-particle total flux for impinging particle of energy ~1k eV. The NEO surface is assumed to consist of CM chondrites.*

***Fig.6:*** *Expected difference (density(CI)- density (CM)) in the intensity of the sputtered-H density emerging from a NEO surface. For a* CI chondrite the expected *sputtered-H density* is slightly bigger than that for a CM chondrite.

***Fig.7****: Different projections of the RAMON basic concept*

*The RAMON sensor consists of the following subsystems:*

*A: Cover (not shown); B: Parallel plate collimator, balanced biased +5kV –5kV;*

*GASP: C1: nanotube for ionizing lower energies particles, C2: electronic gate, C3: ESA, C4: MCP, C5: 2D Anode system (not shown).*

*The particle (red line) enters from the left side (B), gets ionized (blue line) passing through C1, accelerated by C2, deflected by C3, and finally detected by C4.*

*SHEAMON: D1 two nanogrids and the shuttering system, D2: Conversion Surface; D3 MCP electron detector; D4: MCP ion detector; D5: 2D Anode system (not shown).*

*The particle (red line) enters from the left side (B), passes through D1, hits D2, releasing electrons (yellow line) detected by D3. The particle is deviated and ionized (blue line). Finally, the ion is detected by D4.*

**TABLES**

**Table-I**

| **Mass (amu)** | **Element** | **CI (atoms%)** | **CM (atoms%)** | **Tagish Lake (atoms%)** |
|---|---|---|---|---|
| 1 | H | 55 | 45 | 47 |
| 12 | C | 8 | 6 | 9 |
| 24 | Mg | 11 | 15 | 14 |
| 27 | Al | 1 | 1 | 1 |
| 28 | Si | 10 | 15 | 13 |
| 32 | S | 4 | 3 | 3 |
| 40 | Ca | 1 | 1 | 1 |
| 56 | Fe | 9 | 13 | 11 |
| 59 | Ni | 1 | 1 | 1 |
| | Total | 100 | 100 | 100 |

**Table-II**

| Parameter name | Symbol | Suggested Value |
|---|---|---|
| Solar-wind flux | $\varphi_{H+}$ | $10^{12}$ $m^{-2}s^{-1}$ |
| Energy of the incident particle | $E_i$ | 1000 eV |
| Mass of the incident particle | $m_1$ | 1 AMU (proton) |
| NEO Radius | $R_{NEO}$ | 500 m |
| NEO Mass | $M_{NEO}$ | $10^{12}$ kg |
| Mass of the ejected particle | $m_2$ | See Table-I |
| Sputtering Yield | Y | see text |
| Binding energy | $E_b$ | 2 eV |
| Exponent | $k$ | 1 |

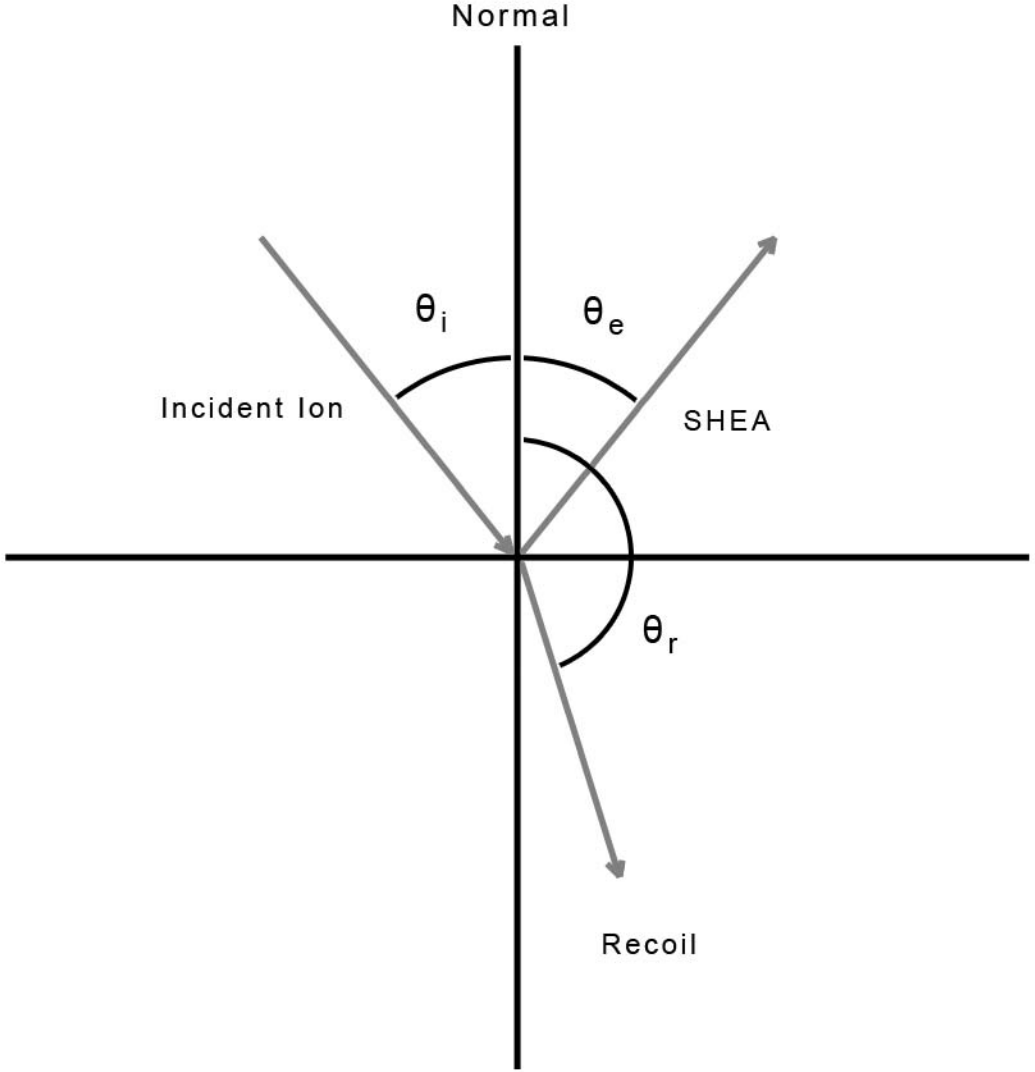

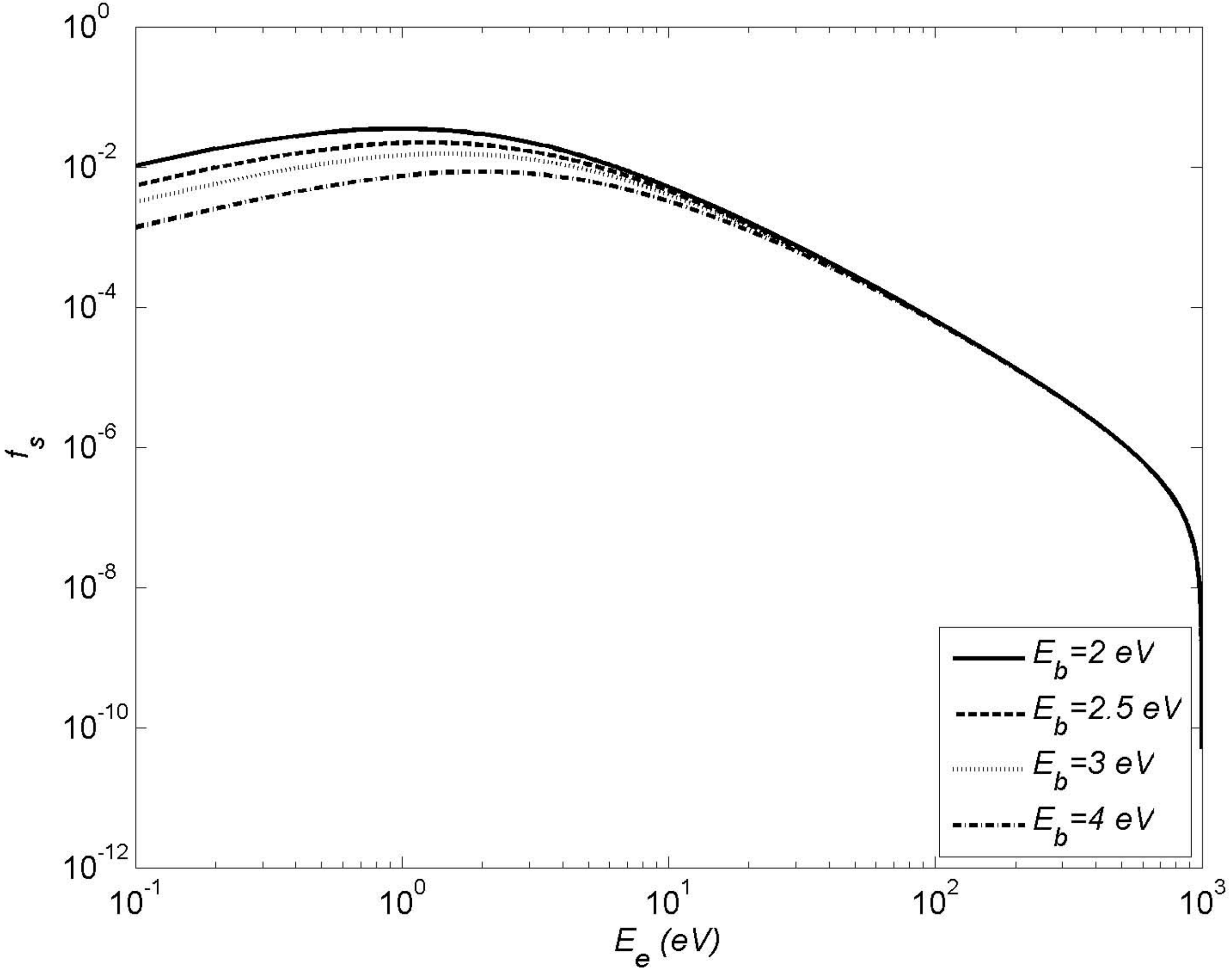

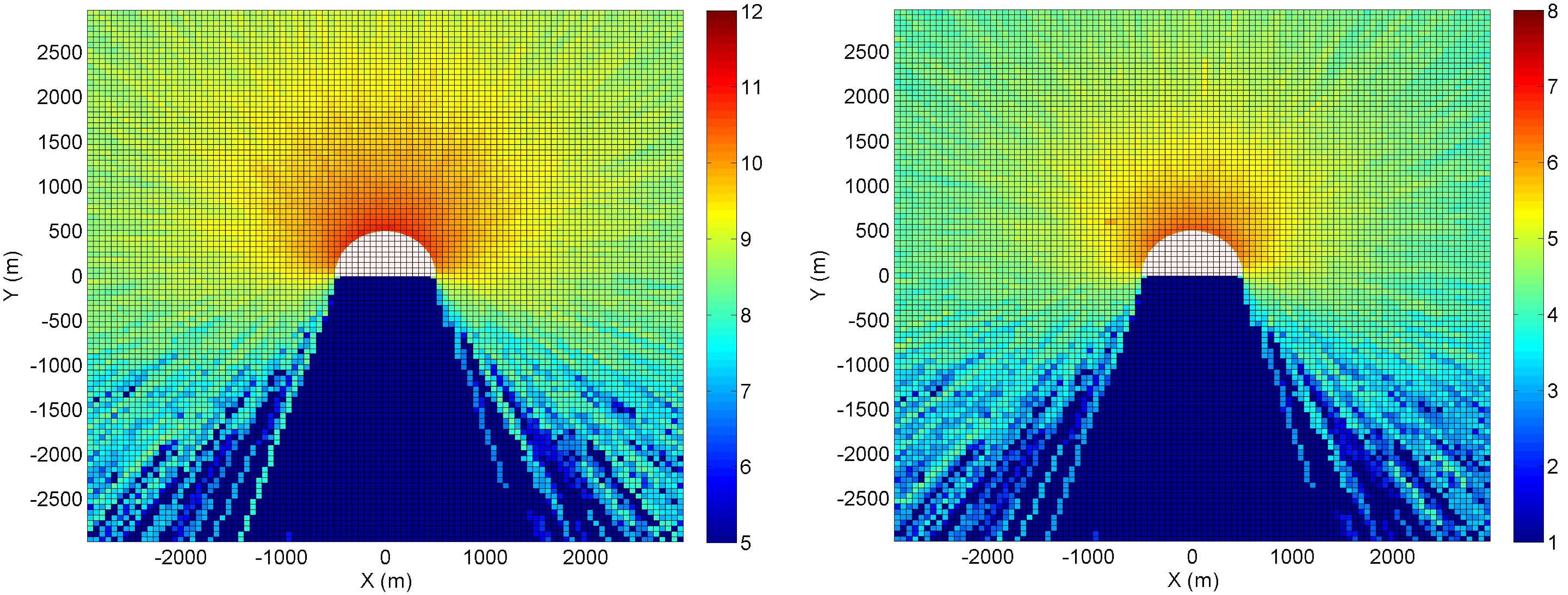

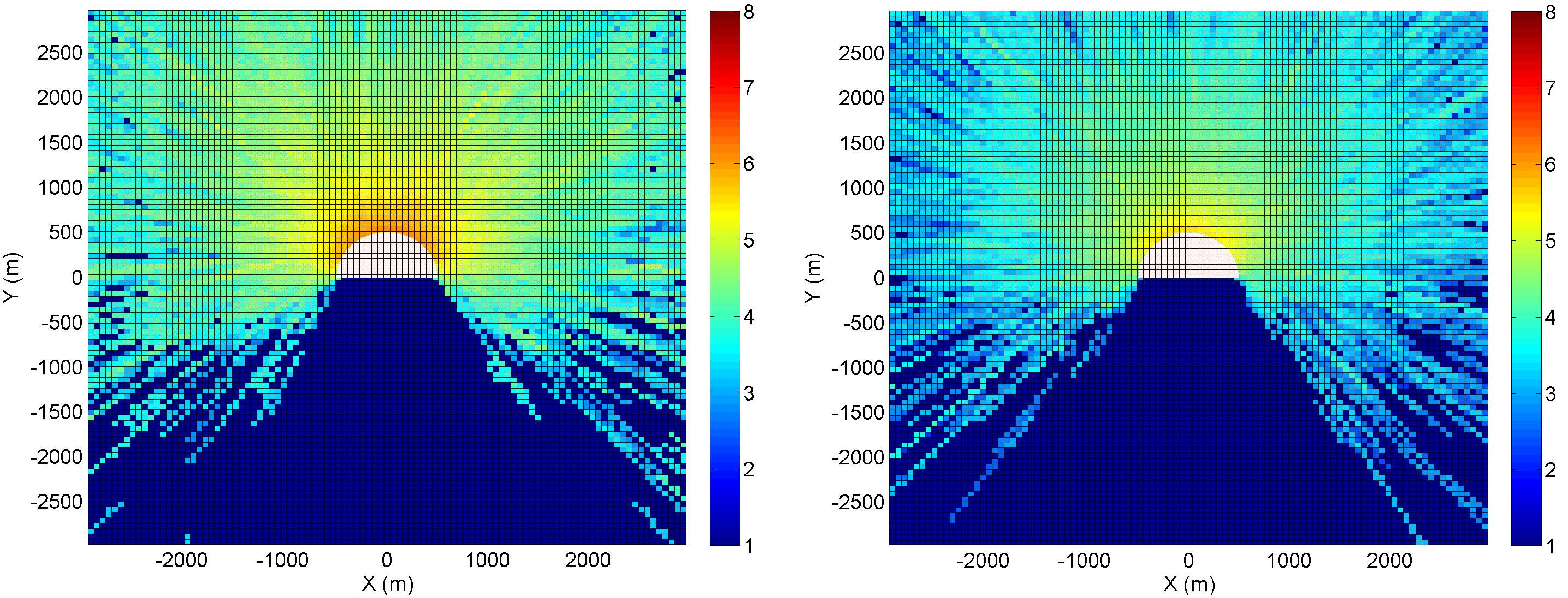

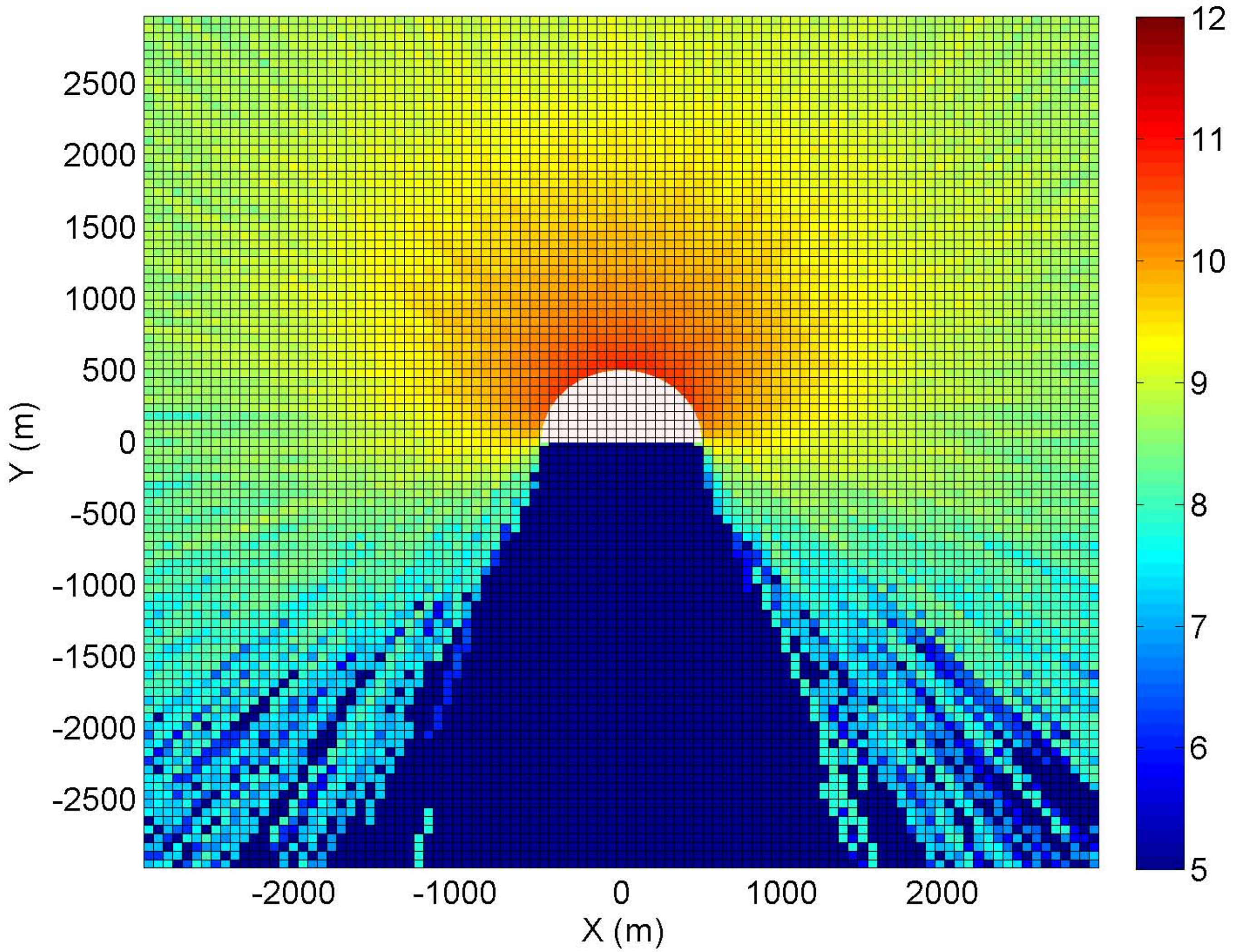

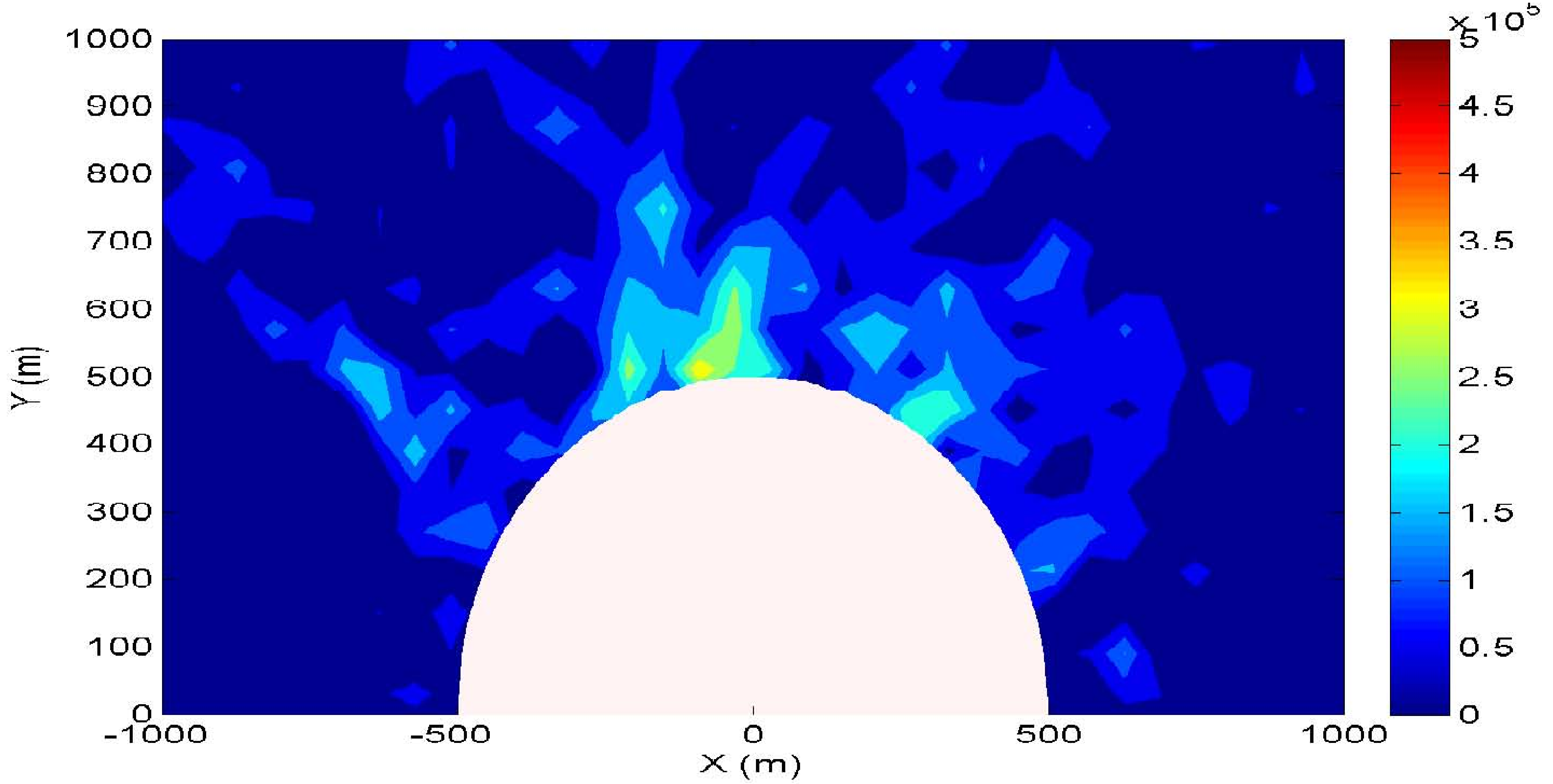

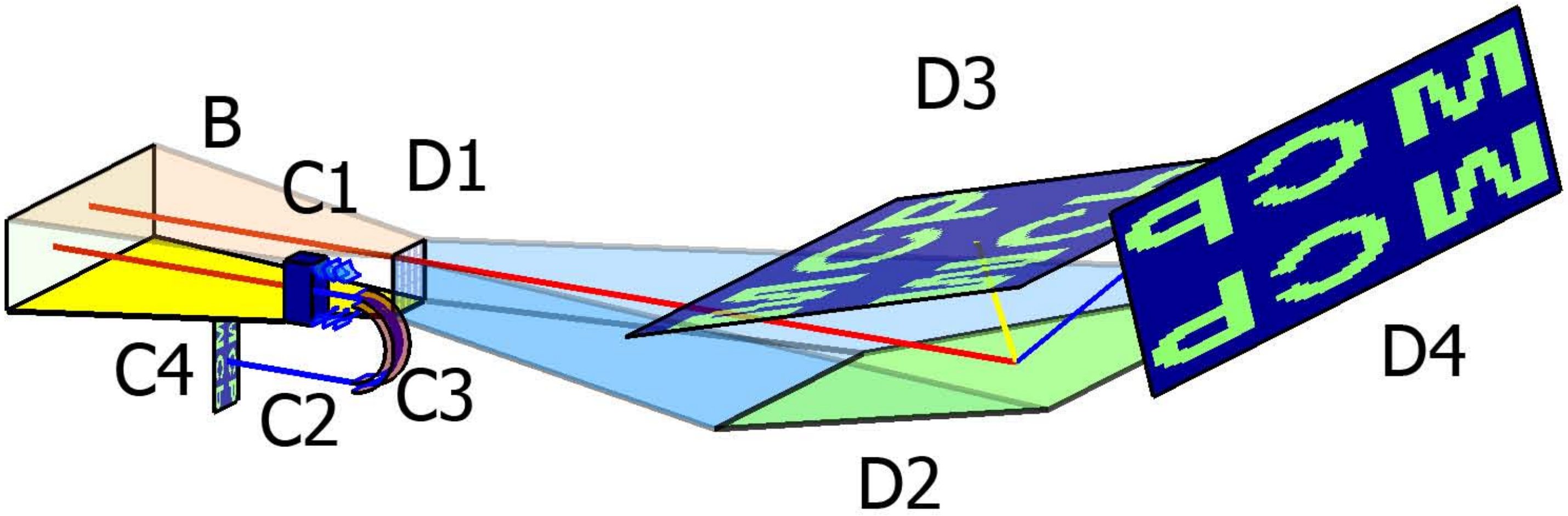

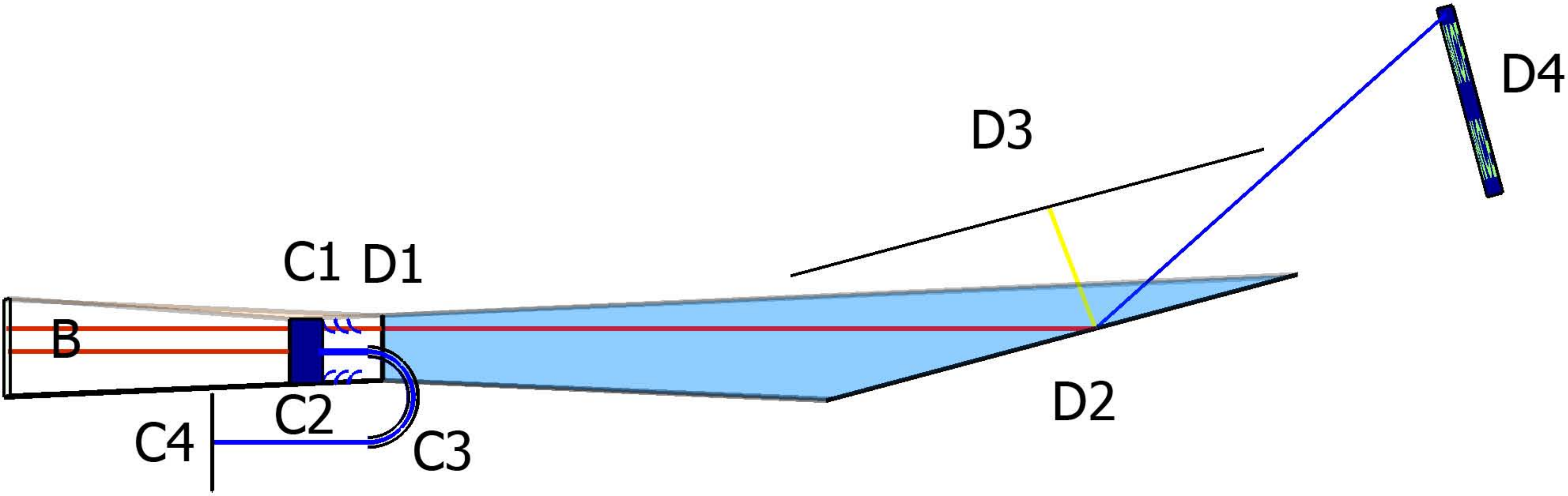